\documentclass[aps,prfluids,final,superscriptaddress,tightenlines,11pt]{revtex4-2}

\newcommand{\Order}[1]{O(#1)}

\usepackage{mathtools} % loads amsmath as well
\usepackage{amssymb}
\usepackage{bm}
\usepackage{graphicx}
\usepackage{color}
\graphicspath{{figs/}}
\usepackage{hyperref}
\usepackage{subcaption}
\usepackage{float}

\let\oldleft\left
\let\oldright\right
\renewcommand{\left}{\mathopen{}\mathclose\bgroup\oldleft}
\renewcommand{\right}{\aftergroup\egroup\oldright}
\renewcommand{\l}{\left}
\renewcommand{\r}{\right}

\newcommand{\ee}{\mathrm{e}}

\newcommand{\eps}{\varepsilon}
\newcommand{\crit}{{\mathrm{crit}}}

\newcommand{\change}[1]{#1}

\begin{document}

%\preprint{}

\title{Reaching for the surface: spheroidal microswimmers in surface gravity waves}


\author{Kunlin Ma}
%\thanks{}
\affiliation{Department of Civil and Environmental Engineering, University of Wisconsin--Madison, Madison WI 53706, USA}
\affiliation{Department of Mathematics, University of Wisconsin--Madison, Madison, WI 53706, USA}

\author{Nimish Pujara}
\email[]{npujara@wisc.edu}
%\thanks{}
\affiliation{Department of Civil and Environmental Engineering, University of Wisconsin--Madison, Madison WI 53706, USA}

\author{Jean-Luc Thiffeault}
%\email[]{npujara@wisc.edu}
%\thanks{}
\affiliation{Department of Mathematics, University of Wisconsin--Madison, Madison, WI 53706, USA}

\date{\today}

\begin{abstract}
Microswimmers (planktonic microorganisms or artificial active particles) immersed in a fluid interact with the ambient flow, altering their trajectories. In surface gravity waves, a common goal for microswimmers is vertical migration (\textit{e.g.}, to reach the free surface or to dive to deeper depths). By modelling microswimmers as spheroidal bodies with an intrinsic swimming velocity that supplements advection and reorientation by the flow, we investigate how shape and swimming affect vertical transport of microswimmers in waves. We find that it is possible for microswimmers to be initially swimming downwards, but to recover and head back to the surface, and vice versa. This is because the coupling between swimming and flow-induced reorientations introduces a shape dependency in the vertical transport. From a wave-averaged analysis of microswimmer trajectories, we show that each trajectory is bounded by critical planes in the position-orientation phase space that depend only on the shape. We also give explicit solutions to these trajectories and determine the fraction of microswimmers that begin within the water column and eventually reach the surface. For microswimmers that are initially randomly oriented, the fraction that reach the surface increases monotonically as the starting depth decreases, as expected, but also varies with shape and swimming speed. In the limit of small swimming speed, the \change{fraction of highly prolate microswimmers reaching the surface is 0.5, suggesting that these swimmers would be able to choose direction of vertical transport with small changes in swimming behaviour.}
\end{abstract}

%\keywords{}

\maketitle

\section{Introduction\label{sec:intro}}

Interactions between a microswimmer and the background flow field control its transport. For example, in vortex lattice flows, microswimmers can become concentrated at the edges of vortices, and potentially escape, depending on their shape and swimming speed \citep{Torney07}, while variations in swimming speed alters diffusivity for spherical swimmers \citep{Khurana11, Khurana12} and induces wildly oscillating transport properties for oblate swimmers \citep{Berman20, Berman21}. Similarly, within a single vortex, shape and swimming determine whether microswimmers remain trapped \citep{ArguedasLeiva20} or experience rapid ejection \citep{Sokolov16}. In narrow channels, microswimmer transport depends on the relative flow strength and interactions with walls, which are different for pushers and pullers \citep{ZottlStark12}. Effects of shape and swimming also persist in turbulent and shear flows, where microswimmers can aggregate due to preferential alignments with the velocity field and local velocity gradients \citep{Zhan13, Pujara18, Borgnino19}, due to gyrotactic trapping \citep{Durham09, Santamaria14, Borgnino18, Cencini19}, and due to interactions between microswimmers \citep{Breier18}.

\begin{figure}
\includegraphics[width=0.75\textwidth]{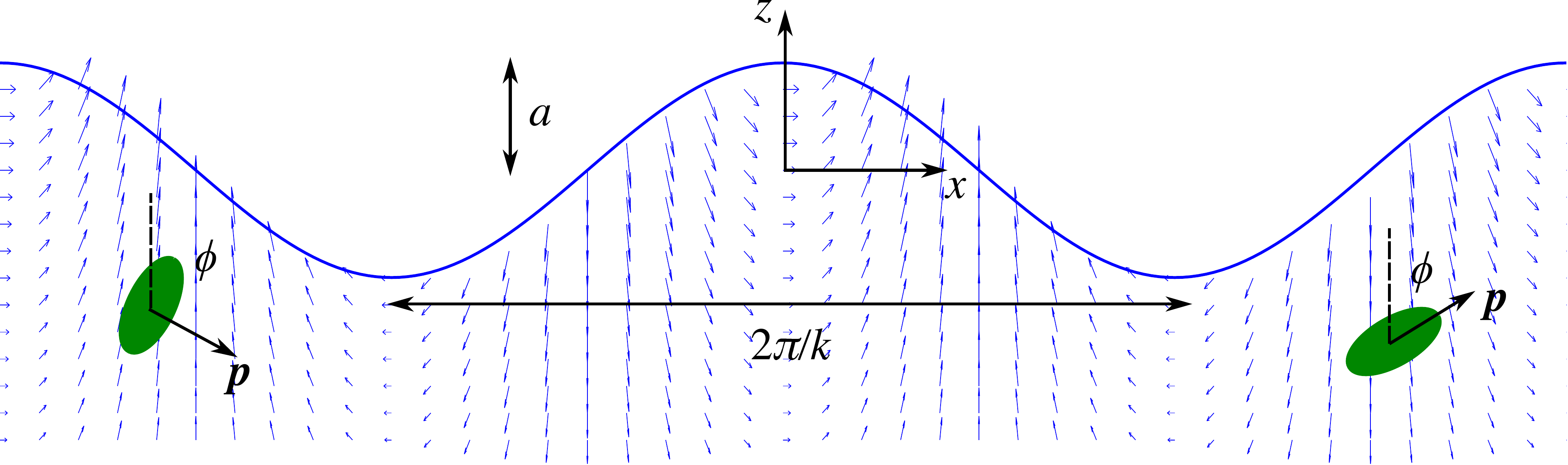}
\caption{Definition sketch of axisymmetric ellipsoidal microswimmers in surface gravity waves swimming along their axis of symmetry (left oblate and right prolate).\label{fig:definition_sketch}}
\end{figure}

In this paper, we examine microswimmer transport within the flow field induced by surface gravity waves. This work is motivated by microswimmers in nature (\textit{e.g.}, motile plankton) navigating flow near the surface of oceans and lakes that includes waves. In reality, the ocean surface will be a complex three-dimensional mixture of waves and turbulence with motile plankton that have complex shapes, mass distributions, and swimming behavior. However, as a starting point we consider here a simplified scenario that captures the essence of how microswimmer transport is controlled by its interactions with the wave-induced flow. The microswimmers are modelled as neutrally buoyant spheroids that swim with an intrinsic velocity along their axis of symmetry, which remains co-planar within the flow field induced by two-dimensional linear waves (Fig.~\ref{fig:definition_sketch}). Our approach follows previous work on active particles in vortex flows \citep{Torney07, Khurana11, Khurana12, Berman20, Berman21, ArguedasLeiva20, Sokolov16, Zhan13, Pujara18, Borgnino19} and passive particles in surface waves \citep{Eames08, Santamaria13, BakhodayPaskyabi15, DiBenedetto18a, DiBenedetto18b} to consider active particles in a wave-induced flow field.

We first show that microswimmer trajectories can be described by a set of coupled ordinary differential equations (Section~\ref{sec:model_equation}) from which the wave-averaged microswimmer trajectories can be found using a two-timescale expansion (Section~\ref{sec:twotime}). We use these to identify regions of phase space in which microswimmers begin swimming downwards but return to the surface and we calculate the probability of this return (Section~\ref{sec:trajectories}). Finally, we discuss our results in the context of more realistic scenarios \change{(Section~\ref{sec:discuss} and Appendix~\ref{sec:3D})}.

\section{Model equations}\label{sec:model_equation}
Small-amplitude progressive surface gravity waves travelling in the $x$ direction in deep water are described by
\begin{subequations}
\label{eq:linear_waves}
\begin{align}
  \eta &= a \cos (kx - \omega t) \label{eq:linear_waves1} \\
  u_x &= a \omega \, \ee^{kz} \cos (kx - \omega t) \label{eq:linear_waves2} \\
  u_z &= a \omega \, \ee^{kz} \sin (kx - \omega t). \label{eq:linear_waves3}
\end{align}
\end{subequations}
Here, $z = \eta$ is the free-surface position, $a$ is the wave amplitude, $k$ is the wavenumber, $\omega$ is the angular frequency, $\bm{u} = (u_x, u_z)$ gives the fluid velocity field, and ${\omega}^2 = gk$ is the dispersion relation.

In this flow, the motion of small, spheroidal microswimmers with intrinsic swimming velocity~$V_{\mathrm{s}}$ is given by
\begin{subequations}
\label{eq:microswimmers}
\begin{align}
    {\bm{v}} &= {\bm{u}} + V_{\mathrm{s}}\,{\bm{p}} \label{eq:microswimmers1} \\
    \dot{\bm{p}} &= \bm{\Omega}\,\bm{p} + \lambda \left[\bm{S}\bm{p} - ({\bm{p}}^{T}\bm{S}\,\bm{p})\bm{p} \right]. \label{eq:microswimmers2}
\end{align}
\end{subequations}
The microswimmer velocity $\bm{v}$ is the vector sum of the fluid velocity and its swimming velocity with $\bm{p}$ being a unit vector that points along the microswimmer's direction of swimming. The rate of change of microswimmer orientation $\dot{\bm{p}} = \mathrm{d}\bm{p}/ \mathrm{d}t$ is given by \citeauthor{Jeffery22}'s (\citeyear{Jeffery22}) equation, where $\bm{\Omega} = \tfrac12[\nabla \bm{u} - (\nabla \bm{u})^T]$ and $\bm{S} = \tfrac12[\nabla \bm{u} + (\nabla \bm{u})^T]$ are the local rotation rate and strain rate tensors, respectively. The particle shape is described by the aspect ratio of the body $\mathrm{AR}$, which is defined to be the ratio of the diameter parallel to the axis of symmetry to the diameter perpendicular to the axis of symmetry. The aspect ratio enters the problem via the eccentricity $\lambda = ({\mathrm{AR}^2 - 1})/({\mathrm{AR}^2 + 1})$, which takes values between~$-1$ and~$1$, with positive values indicating prolate spheroids (rod-shaped) and negative values indicating oblate spheroids (disk-shaped).

Since the flow is irrotational ($\bm{\Omega} \equiv 0$), the first term in Eq.~\eqref{eq:microswimmers2} is identically zero. Further, the strain rate tensor is described by its two unique components $S_{xx} = -S_{zz} = -ka\omega\ee^{kz}\sin{(kx-\omega t)}$ and $S_{xz} = S_{zx} = ka\omega\ee^{kz}\cos{(kx-\omega t)}$, and with microswimmer motion restricted to be co-planar with the flow, the components of Eq.~\eqref{eq:microswimmers2} are given by
\begin{subequations}
\label{eq:Jefferys}
\begin{align}
    \lambda^{-1}\,\dot{p}_x &= S_{xx}p_{x}(1-(p_{x}^2-p_{z}^2))+S_{xz}p_{z}(1-2p_{x}^2) \\
    \lambda^{-1}\,\dot{p}_z &=
    S_{xz}p_{x}(1-2p_{z}^2)-S_{xx}p_{z}(1+(p_{x}^2-p_{z}^2)).
\end{align}
\end{subequations}
We can use the polar angle $\phi$ to define the swimming direction
\begin{equation}
  p_{x} = \sin\phi, \qquad p_{z} = \cos\phi,
  \label{eq:swimdir}
\end{equation}
where~$\phi=0$ corresponds to swimming upwards against the direction of gravity ($+z$ direction), and~$\phi=\pi/2$ in the direction of wave propagation ($+x$ direction).  Equation~\eqref{eq:Jefferys} can then be reduced to a single equation for $\phi$:
\begin{equation}
\label{eq:Jefferys_reduced}
    \lambda^{-1}\dot{\phi} = S_{xx}\sin{2\phi}+S_{xz}\cos{2\phi}.
\end{equation}

We now have a dynamical system with 3 dependent variables:
\begin{subequations}
\label{eq:ODEs_dimensional}
\begin{align}
    \dot{x} &= a\omega\,\ee^{kz}\cos{(kx-\omega t)}+V_{\mathrm{s}}\sin\phi\\
    \dot{z} &= a\omega\,\ee^{kz}\sin{(kx-\omega t)}+V_{\mathrm{s}}\cos{\phi}\\
    \dot{\phi} &= \lambda k a\omega\,\ee^{kz} \bigl[\cos{(kx-\omega t)\cos 2\phi} - \sin{(kx-\omega t)}\sin 2\phi  \bigr]
\end{align}
\end{subequations}
where $(x,z)$ is the microswimmer position and $\phi$ is the swimming direction

The system described by Eq.~\eqref{eq:ODEs_dimensional} can be made dimensionless by defining $ t' = \omega t$, $\bm{x}' = k \bm{x}$, where we immediately drop the primes, to give
\begin{subequations}
\label{eq:ODEs_dimensionless}
\begin{align}
    \dot{x} &= \alpha\,\ee^{z}\cos{(x - t)}
    + \nu\sin\phi \\
    \dot{z} &= \alpha\,\ee^{z}\sin{(x - t)}
    + \nu\cos{\phi}\\
    \dot{\phi} &= \lambda \alpha\,\ee^{z}\,
    \cos{(x - t + 2\phi)}.
    \label{eq:dphi_dimensionless}
\end{align}
\end{subequations}
All variables are dimensionless from this point forth. The dimensionless groups that characterise this system are the wave steepness $\alpha = ka$, the dimensionless swimming speed $\nu = kV_{\mathrm{s}}/\omega$, and microswimmer shape eccentricity $\lambda = ({\mathrm{AR}^2 - 1})/({\mathrm{AR}^2 + 1})$. Our choices for flow and microswimmer models place certain restrictions on the magnitudes of these dimensionless numbers. The flow model (Eq.~\eqref{eq:linear_waves}) assumes small wave steepness (\textit{i.e.}, $\alpha \ll 1$), but it is commonly used in modelling particle transport in the ocean and its accuracy is well-established for $\alpha \le \Order{10^{-1}}$ \citep{DeanDalrymple91, vandenBremer17}. In the microswimmer model (Eq.~\eqref{eq:microswimmers}), we can take~$\nu \ge 0$ without loss of generality since the microswimmer orientation is captured by~$\phi$. The model assumes that the swimmer is small and swims at a speed that is small compared to the characteristic flow velocity so that the microswimmer motion relative to the fluid is in the inertialess limit.  Typical sizes and swimming speeds of motile oceanic plankton (\textit{e.g.}, dinoflagellates, ciliates, larvae) fall in the range $10^{-5}$--$10^{-3}$ m and $10^{-4}$--$10^{-3}$ m/s, respectively \citep{FuchsGerbi16}\change{, whereas typical fluid velocities associated with surface waves in the ocean  fall in the range $10^{-1}$--$10^{0}$ m/s and typical wave celerities fall in the range $10^{0}$--$10^{1}$ m/s}. Thus, typical swimming speeds are at least an order of magnitude smaller than wave-induced velocities (\textit{i.e.}, $\nu \ll \alpha$) and reasonable dimensionless swimming speeds are given by $\nu \le O(10^{-2})$. Finally, the eccentricity can take any value in the range~$\lambda \in [-1,1]$, which covers the full range of spheroidal shapes \change{from the highly oblate (disk-like) to the highly prolate (fiber-like)}.

\section {Two-timescale expansion\label{sec:twotime}}
\begin{figure}
  \begin{center}
  \includegraphics[width=.55\textwidth]{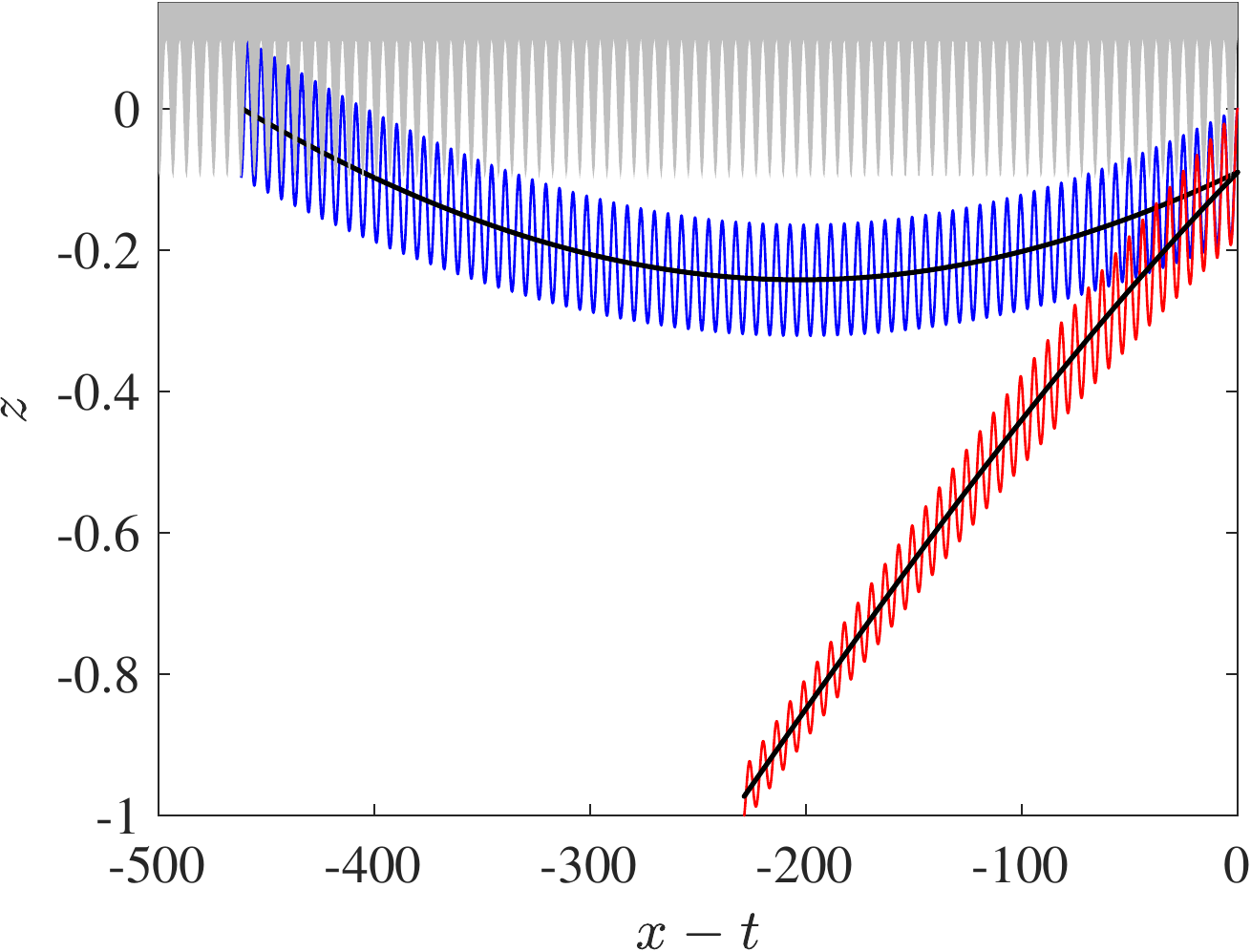}
  \end{center}
  \caption{Two solutions to Eqs.~\eqref{eq:ODEs_dimensionless} (blue and
    red) and to the averaged
    Eqs.~\eqref{eq:ODEs_dimensionless_eps2_solv} (black) for
    $\alpha=0.1$, $\nu=0.005$, $\lambda=0.6$, $x(0)=z(0)=0$, where the initial conditions are adjusted according
    to Eq.~\eqref{eq:IC_avg}. The two initial
    angles are~$\phi(0) = \phi_\crit \mp 0.2$,
    where~$\phi_\crit \approx 2.0895$ is given by Eq.~\eqref{eq:Phicrit} adjusted according
    to Eq.~\eqref{eq:phi0}.  The gray shaded area at the top is the free surface.}
  \label{fig:2traj_average_Xz}
\end{figure}

Sample trajectories from numerical simulations of Eqs.~\eqref{eq:ODEs_dimensionless} in Fig.~\ref{fig:2traj_average_Xz} (computed using \textit{ode45} in MATLAB) show that a microswimmer that is initially near the free surface can end up back at the free surface, or continue swimming to infinite depth depending upon small changes in its initial orientation. The trajectory is shown in a coordinate system moving with the waves, so that the microswimmer appears to be travelling backwards in a frozen wave.
The trajectories in Fig.~\ref{fig:2traj_average_Xz} indicate that the solutions to Eqs.~\eqref{eq:ODEs_dimensionless} consist of fast oscillations at the surface wavelength ($2\pi$ in dimensionless units) superposed with a slower trend at a longer timescale proportional to~$1/\nu$. This suggests using a multiple timescale expansion to remove the fast oscillations.

\subsection {Wave-averaged motion}
We rewrite the solution vector~$\bm{V} = (x\ z\ \phi)^T$ in terms of two timescales:
\begin{equation}
  \bm{V}(t) = \bm{V}^\eps(t,T),
  \qquad
  T = \eps^2\,t,
  \label{eq:twotime}
\end{equation}
where the dependence on the fast timescale~$t$ is assumed periodic.  Since~$\alpha$ and~$\nu$ are small, we express them in terms of a small parameter as~$\alpha \rightarrow \eps\,\alpha$, $\nu \rightarrow \eps^2\,\nu$, with $\nu$ being an order smaller than $\alpha$ as discussed above. Equation~\eqref{eq:ODEs_dimensionless} is then
\begin{subequations}
\label{eq:ODEs_dimensionless_eps}
\begin{align}
  \partial_t{x^\eps} + \eps^2\,\partial_T x^\eps
  &= \eps\alpha\,\ee^{z^\eps}\cos{(x^\eps - t)}
  + \eps^2\nu\sin\phi^\eps \\
  \partial_t{z^\eps} + \eps^2\,\partial_T z^\eps
  &= \eps\alpha\,\ee^{z^\eps}\sin{(x^\eps - t)}
  + \eps^2\nu\cos{\phi^\eps}\\
  \partial_t{\phi^\eps} + \eps^2\,\partial_T \phi^\eps
  &= \eps\lambda\alpha\,\ee^{z^\eps} \cos{(x^\eps - t + 2\phi^\eps)}.
\end{align}
\end{subequations}
We now expand the solution vector in the usual manner:
\begin{equation}
  \bm{V}^\eps(t,T)
  =
  \bm{V}_0(t,T) + \eps\,\bm{V}_1(t,T) + \eps^2\,\bm{V}_2(t,T) + \ldots.
  \label{eq:Vexp}
\end{equation}
At order~$\eps^0$, Eq.~\eqref{eq:ODEs_dimensionless_eps} is
simply~$\partial_t{x_0} = \partial_t{z_0} = \partial_t{\phi_0} = 0$, which indicates that the leading-order quantities are only a function of the slow time~$T$,
\begin{equation}
  x_0 = X(T), \qquad z_0 = Z(T),\qquad \phi_0 = \Phi(T).
  \label{eq:slowvar}
\end{equation}
At order~$\eps^1$, Eq.~\eqref{eq:ODEs_dimensionless_eps} is
\begin{subequations}
\label{eq:ODEs_dimensionless_eps1}
\begin{align}
  \partial_t{x_1}
  &= \alpha\,\ee^{Z}\cos{(X - t)} \\
  \partial_t{z_1}
  &= \alpha\,\ee^{Z}\sin{(X - t)} \\
  \partial_t{\phi_1}
  &= \lambda \alpha\,\ee^{Z}\,\cos{(X - t + 2\Phi)}.
\end{align}
\end{subequations}
Notice that the integral from~$[0,2\pi]$ of each right-hand side in Eq.~\eqref{eq:ODEs_dimensionless_eps1} vanishes, which is the solvability condition at this order.  The unique mean-zero solution to Eq.~\eqref{eq:ODEs_dimensionless_eps1} is
\begin{subequations}
\label{eq:ODEs_dimensionless_eps1_sol}
\begin{align}
  x_1
  &= -\alpha\,\ee^{Z}\sin{(X - t)} \\
  z_1
  &= \alpha\,\ee^{Z}\cos{(X - t)} \\
  \phi_1
  &= -\lambda \alpha\,\ee^{Z}\sin{(X - t + 2\Phi)}.
\end{align}
\end{subequations}

At order~$\eps^2$, Eq.~\eqref{eq:ODEs_dimensionless_eps} is
\begin{subequations}
\label{eq:ODEs_dimensionless_eps2}
\begin{align}
  \partial_t{x_2} + \partial_T X
  &= \alpha\,\ee^{Z}
  \left(\cos{(X - t)\,z_1 - \sin(X - t)\,x_1}\right)
  + \nu\sin\Phi \\
  \partial_t{z_2} + \partial_T Z
  &= \alpha\,\ee^{Z}
  \left(\sin{(X - t)\,z_1 + \cos(X - t)\,x_1}\right)
  + \nu\cos\Phi \\
  \partial_t{\phi_2} + \partial_T \Phi
  &= \lambda \alpha\,\ee^{Z}
  \left(
  \cos(X - t + 2\Phi)\,z_1
  -
  \sin(X - t + 2\Phi)\,(x_1 + 2\phi_1)
  \right).
\end{align}
\end{subequations}
At this order there is a nontrivial solvability condition, obtained by averaging Eq.~\eqref{eq:ODEs_dimensionless_eps2} over a period:
\begin{subequations}
\label{eq:ODEs_dimensionless_eps2_solv}
\begin{align}
  \partial_T X &= \nu\sin\Phi + \alpha^2\,\ee^{2Z}
  \label{eq:ODEs_dimensionless_eps2_solv_dTX} \\
  \partial_T Z  &= \nu\cos\Phi
  \label{eq:ODEs_dimensionless_eps2_solv_dTZ}
  \\
  \partial_T \Phi  &=  \lambda \alpha^2\,\ee^{2Z} \left(\lambda + \cos2\Phi\right).
  \label{eq:ODEs_dimensionless_eps2_solv_dTphi}
\end{align}
\end{subequations}
These are the sought-after governing equations for the slow motion. The final term in Eq.~\eqref{eq:ODEs_dimensionless_eps2_solv_dTX} is the familiar Stokes drift that stems from unclosed wave orbital motions and results in the net transport of fluid particles in the direction of wave propagation \citep{Stokes47}. We shall need explicit solutions to order~$\eps^2$ below, so we substitute the solvability condition Eq.~\eqref{eq:ODEs_dimensionless_eps2_solv} and solution Eq.~\eqref{eq:ODEs_dimensionless_eps1_sol} into Eq.~\eqref{eq:ODEs_dimensionless_eps2}, and obtain the simple set of equations
\begin{equation}
  \partial_t{x_2} = 0, \qquad
  \partial_t{z_2} = 0, \qquad
  \partial_t{\phi_2} = -\lambda^2 \alpha^2\,\ee^{2Z}\cos2(X - t + 2\Phi),
  \label{eq:ODEs_dimensionless_eps2_simp2}
\end{equation}
whose unique mean-zero solution is
\begin{equation}
  x_2 = 0,
  \qquad
  z_2 = 0,
  \qquad
  \phi_2 = \tfrac12\lambda^2 \alpha^2\,\ee^{2Z}\sin2(X - t + 2\Phi).
  \label{eq:ODEs_dimensionless_eps2_sol}
\end{equation}

\subsection{Initial conditions for wave-averaged motion \label{sec:IC_avg}}

Figure~\ref{fig:2traj_average_Xz} shows that solutions to Eqs.~\eqref{eq:ODEs_dimensionless} can be quite sensitive to initial conditions \change{(especially near the critical angles discussed in Section~\ref{sec:phicrit})}.  Indeed, observe that the unaveraged trajectories start at~$(0,0)$, whereas the averaged trajectories start below this point.  In order that a solution to the averaged Eqs.~\eqref{eq:ODEs_dimensionless_eps2_solv} properly shadow its corresponding trajectory, the initial conditions for the unaveraged variables must be projected appropriately onto the slow variables, as we describe in this section.

The initial conditions for the full solution vector~$\bm{V}(t)$ are~$\bm{V}(0) = \bm{V}^\eps(0,0) = (x(0),z(0),\phi(0))$.  From the two-timescale expansion Eq.~\eqref{eq:Vexp} and its solutions  (\eqref{eq:slowvar},~\eqref{eq:ODEs_dimensionless_eps1_sol} and ~\eqref{eq:ODEs_dimensionless_eps2_sol}), we have
\begin{subequations}
\begin{align}
  x(0)
  &=
  X(0) - \eps\alpha\,\ee^{Z(0)}\sin{X(0)} + \Order{\eps^3}
  \label{eq:x0}
  \\
  z(0)
  &=
  Z(0) + \eps\alpha\,\ee^{Z(0)}\cos{X(0)} + \Order{\eps^3},
  \label{eq:z0}
  \\
  \phi(0) &=
  \Phi(0)
  - \eps\lambda\alpha\,\ee^{Z(0)}\sin{\left(X(0)+2\Phi(0)\right)}  \nonumber \\
  &\phantom{=}
  + \tfrac12\eps^2\lambda^2 \alpha^2\,\ee^{2Z(0)}\sin2(X(0) + 2\Phi(0))
  + \Order{\eps^3}.
  \label{eq:phi0}
\end{align}
\end{subequations}
We expand~$X(0) = X_0(0) + \eps\,X_1(0) + \ldots$, and similarly for~$Z(0)$ and~$\Phi(0)$, and equate terms at each order to successively solve for the initial conditions for the slow variables in terms of the initial conditions for the unaveraged variables. For instance, at leading order~$X_0(0) = x(0)$, $Z_0(0) = z(0)$, and at order~$\eps$ we have~$X_1(0) = \alpha\,\ee^{z(0)}\sin{x(0)}$, $Z_1(0) = -\alpha\,\ee^{z(0)}\cos{x(0)}$. Following this procedure, we eventually find that given initial conditions~$\bm{V}(0) = (x(0),z(0),\phi(0))$, the initial conditions for the slow variables are
\begin{subequations}
  \label{eq:IC_avg}
  \begin{align}
    X(0) &=
    x(0)
    + \eps\alpha\,\ee^{z(0)}\sin{x(0)} + \Order{\eps^3}
    \label{eq:X0}
    \\
    Z(0) &=
    z(0)
    - \eps\alpha\,\ee^{z(0)}\cos{x(0)}
    + \eps^2\alpha^2\ee^{2z(0)} + \Order{\eps^3}
    \label{eq:Z0}
    \\
    \Phi(0) &= \phi(0) + \eps\lambda\alpha\ee^{z(0)}\sin(x(0) + 2\phi(0))
    \nonumber \\
    &\phantom{=}
    + \tfrac12\eps^2\lambda\alpha^2\ee^{2z(0)}
    \l(
    \lambda\sin 2(x(0) + 2\phi(0))
    - 2\sin2\phi(0)
    \r) + \Order{\eps^3}.
    \label{eq:Phi0}
\end{align}
\end{subequations}
These corrections to the initial condition are small, but they are crucial, particularly when~$\phi(0)$ is near a critical angle~$\phi_\crit$ (Section~\ref{sec:phicrit}).

\section{Analysis of wave-averaged motion}
\label{sec:trajectories}

\subsection{Critical angles and planes}
\label{sec:phicrit}
A striking feature of the wave-averaged equation for $\Phi$ (Eq.~\eqref{eq:ODEs_dimensionless_eps2_solv_dTphi}) is that there exist critical angles~$\Phi_\crit$ such that
\begin{equation}
  \lambda + \cos2\Phi_\crit = 0,
  \label{eq:crit_criterion}
\end{equation}
which implies that~$\Phi = \Phi_\crit$ for all time. (These critical angles are present even without swimming and were previously observed \citep{DiBenedetto18a, DiBenedetto18b}.) We let~$\Lambda = \tfrac12\arccos(-\lambda)$, where~$\arccos(x) \in [0,\pi]$ is the principal branch of~$\cos^{-1}x$, so that~$0 \le \Lambda \le \pi/2$.   The four solutions for the critical angle in Eq.~\eqref{eq:crit_criterion} can be expressed as a vector
\begin{equation}
  \bm{\Phi}_\crit(\lambda)
  =
  (\Lambda_1,\Lambda_2,\Lambda_3,\Lambda_4)
  =
  \l(
  \Lambda ,
  \pi - \Lambda ,
  \pi + \Lambda ,
  2\pi - \Lambda
  \r).
  \label{eq:Phicrit}
\end{equation}
The~$\Lambda_i$ are chosen such that~$\Lambda_i \in [0,2\pi]$ and~$\Lambda_i \le \Lambda_{i+1}$.  The vector of solutions degenerates to
\begin{subequations}
\label{eq:lambdapm1}
\begin{alignat}{2}
  \bm{\Phi}_\crit(-1) &= (0,\pi,\pi,2\pi), \qquad&\text{(disk-shaped microswimmer)} \label{eq:lambdam1} \\
  \bm{\Phi}_\crit(+1) &= (\pi/2,\pi/2,3\pi/2,3\pi/2), \qquad&\text{(fiber-shaped microswimmer)}, \label{eq:lambda1}
\end{alignat}
\end{subequations}
which are the only cases with fewer than four distinct solutions. In the three-dimensional phase space~$(x,z,\phi)$,~$\phi = \Phi_\crit$ are critical planes since the average dynamics cannot cross these planes (at least for $t \lesssim \Order{\eps^{-2}}$). The motion is always confined between two planes and the four planes delimit four invariant regions in the three-dimensional phase space. Note that solutions to the unaveraged system Eq.~\eqref{eq:ODEs_dimensionless} may momentarily cross the critical planes when the average dynamics are near the boundaries due to the fast timescale oscillations.

\subsection{Solutions of wave-averaged trajectories}

\begin{figure}
  \begin{center}
  \subcaptionbox{\label{fig:phaseportrait_Z_Phi_positive}}{%
    \includegraphics[width=0.49\textwidth]{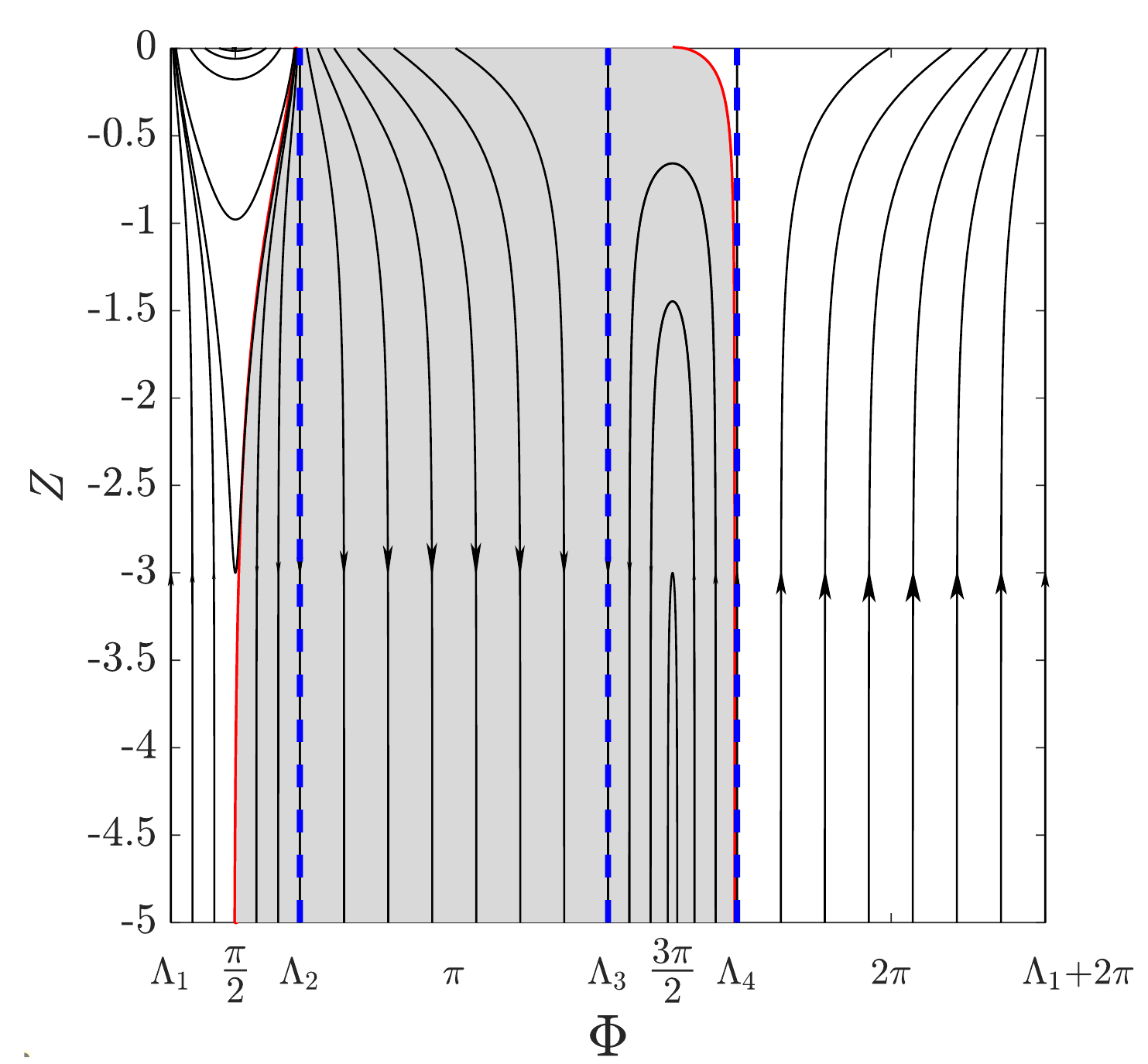}}
  \subcaptionbox{\label{fig:phaseportrait_Z_Phi_negative}}{%
    \includegraphics[width=0.49\textwidth]{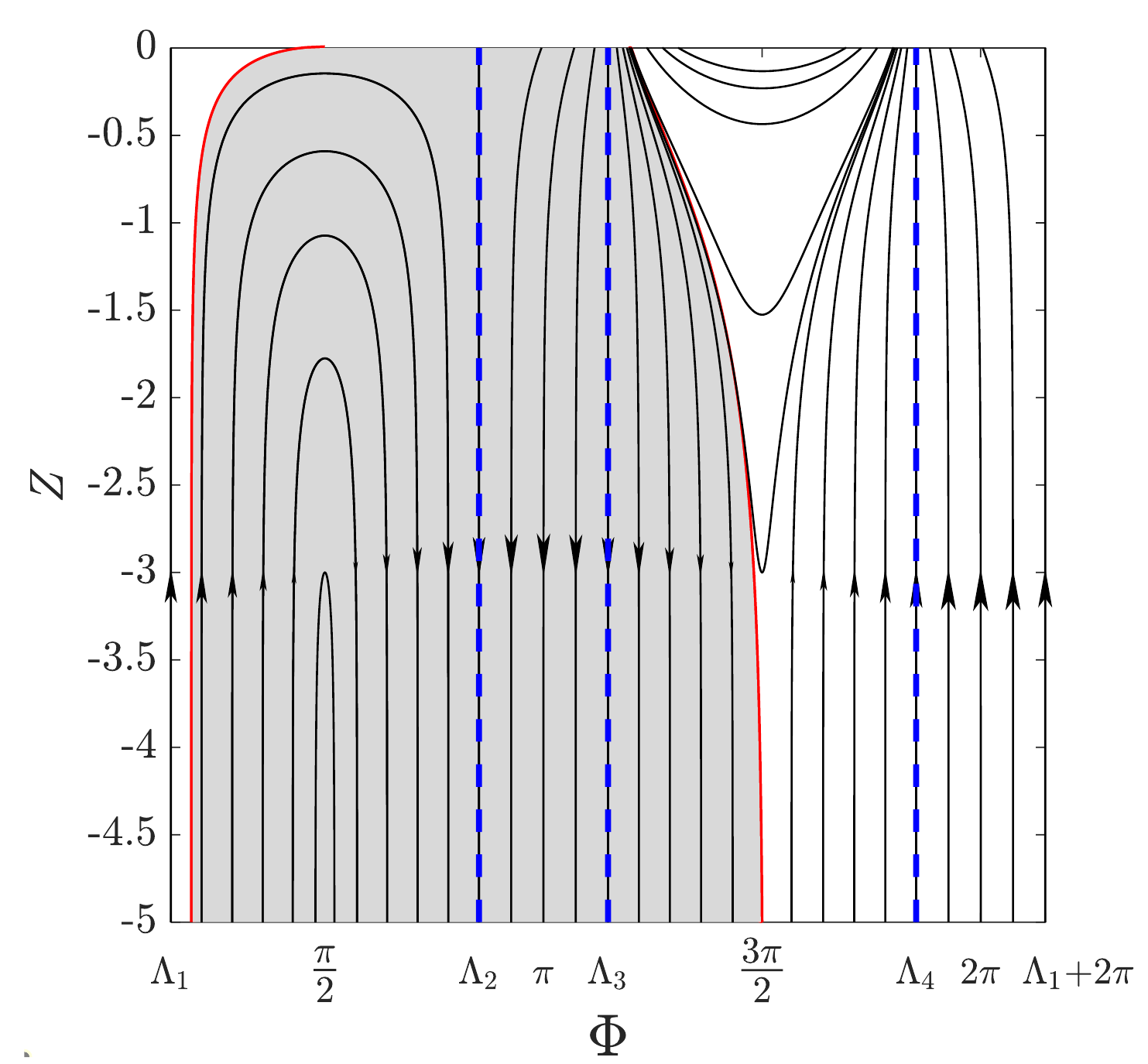}}
  \end{center}
  \caption{$\Phi$--$Z$ phase portrait for
    \eqref{eq:ODEs_dimensionless_eps2_solv} for
    $\alpha=0.1$ and $\nu=0.005$, with (a) $\lambda=0.6$ and (b) $\lambda=-0.6$.  The vertical dashed lines indicate the critical angles~\eqref{eq:Phicrit}.
    The shaded regions corresponds to trajectories that end up at~$Z = -\infty$ and the red curves are the bounding lines of these regions~\eqref{eq:boundingcurve_left_positivelambda}--\eqref{eq:boundingcurve_right_negativelambda}.  }
  \label{fig:phaseportrait}
\end{figure}

Observe that the~$\partial_T Z$ and~$\partial_T \Phi$ equations in Eqs.~\eqref{eq:ODEs_dimensionless_eps2_solv} do not depend on~$X$, and so can be solved separately. It is instructive to plot a phase portrait for the~$\Phi$--$Z$ plane, as in Fig.~\ref{fig:phaseportrait}. The phase portrait shows that swimmer contours do not cross lines where $\Phi = \Phi_\crit$, as discussed in the previous section. It also shows that there are certain regions in this phase space where swimmers reach the surface (unshaded) and other regions where the swimmers dive to infinite depth (shaded gray). Note that swimmers are swimming down in the region $\pi/2 < \Phi < 3\pi/2$.

To find explicit solutions for the contour lines in this phase portrait, we first take the ratio~$\partial_T Z$ over~$\partial_T \Phi$
\begin{equation}
  \frac{d Z}{d\Phi}
  =
  \frac{\nu\ee^{-2Z}\cos{\Phi}}
  {\lambda\alpha^2\left(\cos{2\Phi} - \cos{2\Lambda}\right)}
  \label{eq:dZdPhi}
\end{equation}
where~$\cos2\Lambda = -\lambda$.  This can be rewritten as
\begin{equation}
  \frac{d\left(\ee^{2Z}\right)}{d(\sin{\Phi})}
  =
  \frac{\nu}{\lambda\alpha^2}\,
  \frac{1}{\sin^2{\Lambda} - \sin^2{\Phi}}.
  \label{eq:dexp2ZdsinPhi}
\end{equation}
If the initial condition~$\Phi_0$ ($0$ subscripts here indicate initial conditions, not to be confused with the expansion in Section~\ref{sec:IC_avg}) is such that~$\lvert\sin\Phi_0\rvert = \sin\Lambda$, then $\Phi$ is at a critical angle and is therefore constant according to Eq.~\eqref{eq:Phicrit}. In this case, we cannot express~$Z$ as a function of~$\Phi$.  This is the origin of the singularity in~\eqref{eq:dexp2ZdsinPhi}.  We may thus assume that~$\lvert\sin\Phi_0\rvert \ne \sin\Lambda$. Eq~\eqref{eq:dexp2ZdsinPhi} can then be integrated to give
\begin{equation}
  \ee^{2Z}
  -
  \ee^{2Z_0}
  =
  \frac{\nu}{2\lambda\alpha^2\sin\Lambda}
  \log\l(%\lvert
  \frac{\sin\Lambda + \sin\Phi}{\sin\Lambda - \sin\Phi}\,
  \frac{\sin\Lambda - \sin\Phi_0}{\sin\Lambda + \sin\Phi_0}
  \r)%\rvert
  ,
  \label{eq:exp2Zsol}
\end{equation}
such that~$Z(\Phi_0) = Z_0$.  This solution ceases to exist when~$\lvert\sin\Phi\rvert = \sin\Lambda$: this corresponds to solutions that asymptote to a critical angle.

\subsection{Probability of hitting the surface: effects of shape and swimming speed}
\label{sec:probhit}

For microswimmers with random initial orientation~$\Phi_0$ uniformly distributed in~$[0 \; 2\pi]$ at starting depth $Z_0$, there is a fraction of microswimmers that hit the surface with the rest swimming to infinite depth. This is shown in Fig.~\ref{fig:phaseportrait}, with the gray regions correspond to swimmers that end up at~$Z = -\infty$. We can find the microswimmer fraction hitting the free surface (FHS) by calculating the bounding curves on the left and right sides of the phase portrait (shown as red lines in Fig.~\ref{fig:phaseportrait})
\begin{equation}
    \mathrm{FHS} = 1 - \frac{\Phi_0^{(\text{right})}(Z_0) - \Phi_0^{(\text{left})}(Z_0)}{2\pi}.
    \label{eq:FHS}
\end{equation}
In a given wave field, FHS is a function of shape parameter $\lambda$, swimming speed $\nu$, and starting depth $Z_0$.

To find equations for the trajectories that bound the regions where the swimmers end up at ~$Z = -\infty$, we first let
\begin{equation}
  \Delta(x) :=
  \tanh[x\,\lambda\alpha^2\,\nu^{-1}\sin\Lambda]
  \label{eq:Delta}
\end{equation}
and solve Eq.~\eqref{eq:exp2Zsol} for~$\sin\Phi_0$:
\begin{equation}
  \sin\Phi_0
  =
  \frac{\sin\Phi - \sin\Lambda\,\Delta(\ee^{2Z} - \ee^{2Z_0})}
  {\sin\Lambda - \sin\Phi\,\Delta(\ee^{2Z} - \ee^{2Z_0})}
  \,\sin\Lambda\,.
  \label{eq:sinPhi0sol}
\end{equation}
We consider separately the case of positive and negative~$\lambda$.

\smallskip

\noindent
{\bf Positive ~$\lambda$}: {$0 < \lambda < 1$}

\noindent To find the bounding curves, we consider final values of~$\Phi$ and~$Z$ in Eq.~\eqref{eq:sinPhi0sol}, guided by Fig.~\ref{fig:phaseportrait_Z_Phi_positive}.  We then find~$\Phi_0$ by inverting the sine, taking care to use the appropriate solution branch.  Setting $Z\rightarrow -\infty$ with $\Phi=\pi/2$ in Eq.~\eqref{eq:sinPhi0sol} gives the curve bounding the gray region on the left:
\begin{equation}
  \Phi_0^{(\text{left})}(Z_0)
  =
  \pi -
  \arcsin\l(
  \frac{1 + \sin\Lambda\,\Delta(\ee^{2Z_0})}
  {\sin\Lambda + \Delta(\ee^{2Z_0})}
  \,\sin\Lambda\r)\,.
  \label{eq:boundingcurve_left_positivelambda}
\end{equation}
Setting $Z\rightarrow 0$ with $\Phi=3\pi/2$ in Eq.~\eqref{eq:sinPhi0sol} gives the curve bounding the gray region on the right:
\begin{equation}
  \Phi_0^{(\text{right})}(Z_0)
  =
  2\pi -
  \arcsin\l(
  \frac{1 + \sin\Lambda\,\Delta(1 - \ee^{2Z_0})}
  {\sin\Lambda + \Delta(1 - \ee^{2Z_0})}
  \,\sin\Lambda\r)\,.
  \label{eq:boundingcurve_right_positivelambda}
\end{equation}

\smallskip

\noindent
{\bf Negative ~$\lambda$}: {$-1 < \lambda < 0$}

\noindent We proceed as for~$\lambda>0$, this time guided by Fig.~\ref{fig:phaseportrait_Z_Phi_negative}.  Setting $Z\rightarrow 0$ with $\Phi=\pi/2$ in Eq.~\eqref{eq:sinPhi0sol} gives
\begin{equation}
  \Phi_0^{(\text{left})}(Z_0)
  =
  \arcsin\l(
  \frac{1 - \sin\Lambda\,\Delta(1 - \ee^{2Z_0})}
  {\sin\Lambda - \Delta(1 - \ee^{2Z_0})}
  \,\sin\Lambda\r)\,.
  \label{eq:boundingcurve_left_negativelambda}
\end{equation}
Setting $Z\rightarrow -\infty$ with $\Phi=3\pi/2$ in Eq.~\eqref{eq:sinPhi0sol} gives
\begin{equation}
  \Phi_0^{(\text{right})}(Z_0)
  =
  \pi
  +
  \arcsin\l(
  \frac{1 - \sin\Lambda\,\Delta(\ee^{2Z_0})}
  {\sin\Lambda - \Delta(\ee^{2Z_0})}
  \,\sin\Lambda\r)\,.
  \label{eq:boundingcurve_right_negativelambda}
\end{equation}

\begin{figure}
  \begin{center}
    \includegraphics[width=0.55\textwidth]{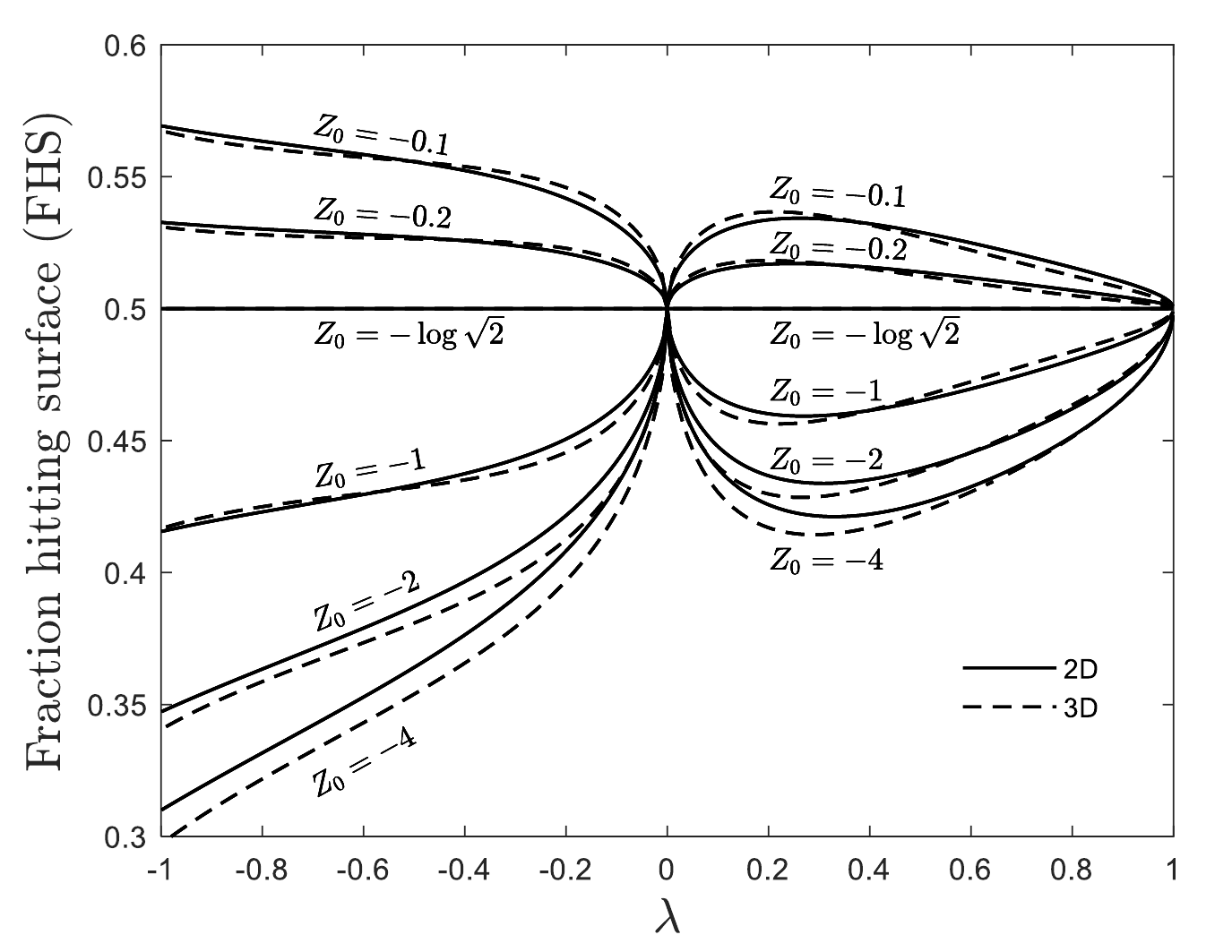}
  \end{center}
  \caption{The fraction (Eq.~\eqref{eq:FHS}) of microswimmers that hit the surface as a function of shape parameter $\lambda$ for different starting depths $Z_0$ and uniformly distributed initial orientation $\Phi_0$ for $\alpha=0.1$ and $\nu=0.005$.  The dashed lines are the similar results for the three-dimensional case (see Appendix~\ref{sec:3D}).}
  \label{fig:fraction_hitting_surface_2D_3D}
\end{figure}

Figure~\ref{fig:fraction_hitting_surface_2D_3D} shows the variation of FHS as a function of shape and starting depth for a fixed swimming speed. FHS increases monotonically as the starting depth $Z_0$ decreases. Strangely, at the special starting depth of $Z_0 = -\log{\sqrt{2}}\approx -0.347$, exactly half of the swimmers hit the surface regardless of their shape. For spherical swimmers ($\lambda = 0$) and fiber-shaped swimmers ($\lambda=1$), exactly half hit the surface regardless of their starting depth.

\begin{figure}
  \begin{center}
  \subcaptionbox{\label{fig:FHS_3depth_nu_0.01}}{%
    \includegraphics[width=0.49\textwidth]{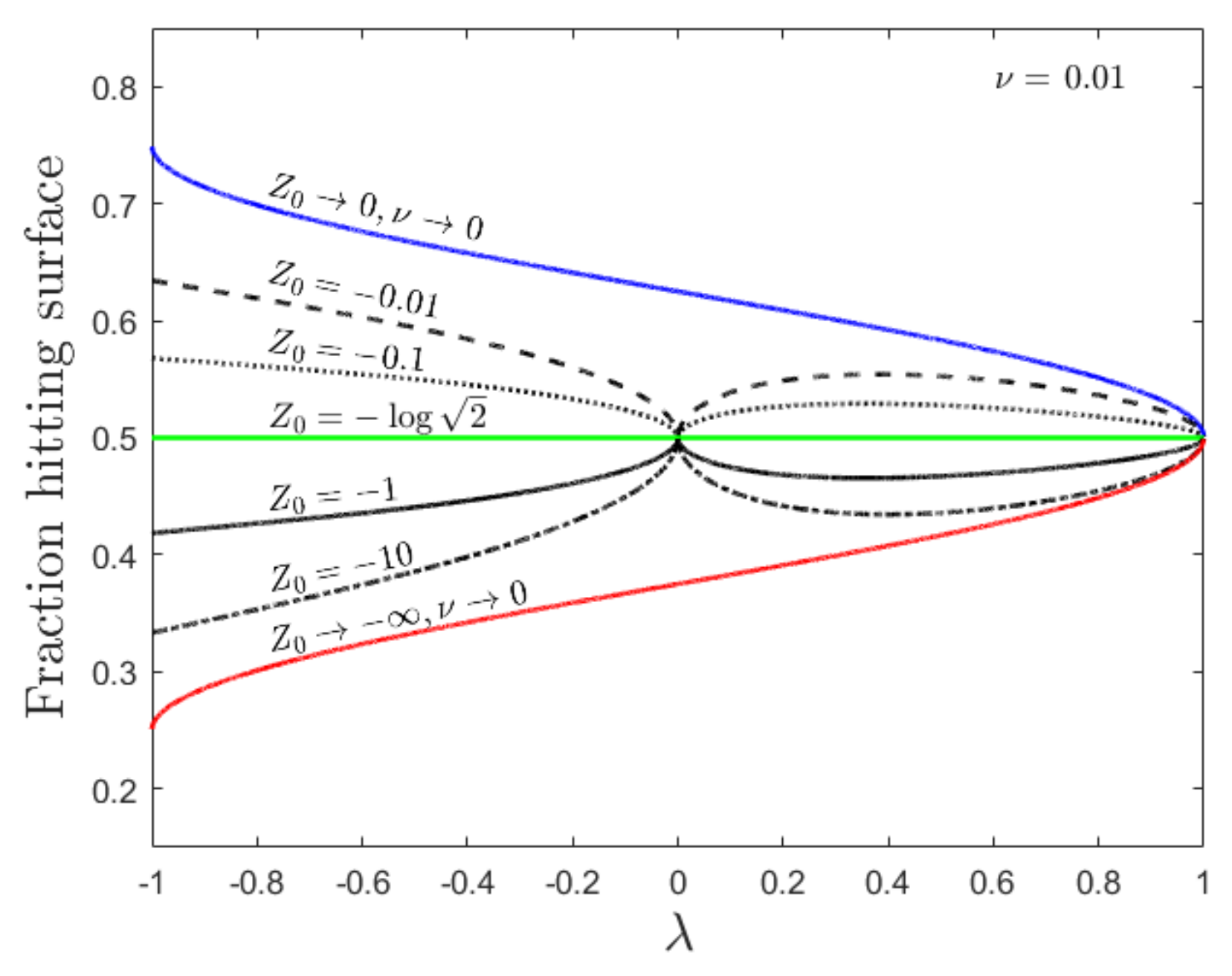}}
  \subcaptionbox{\label{fig:FHS_3depth_nu_0pt005}}{%
    \includegraphics[width=0.49\textwidth]{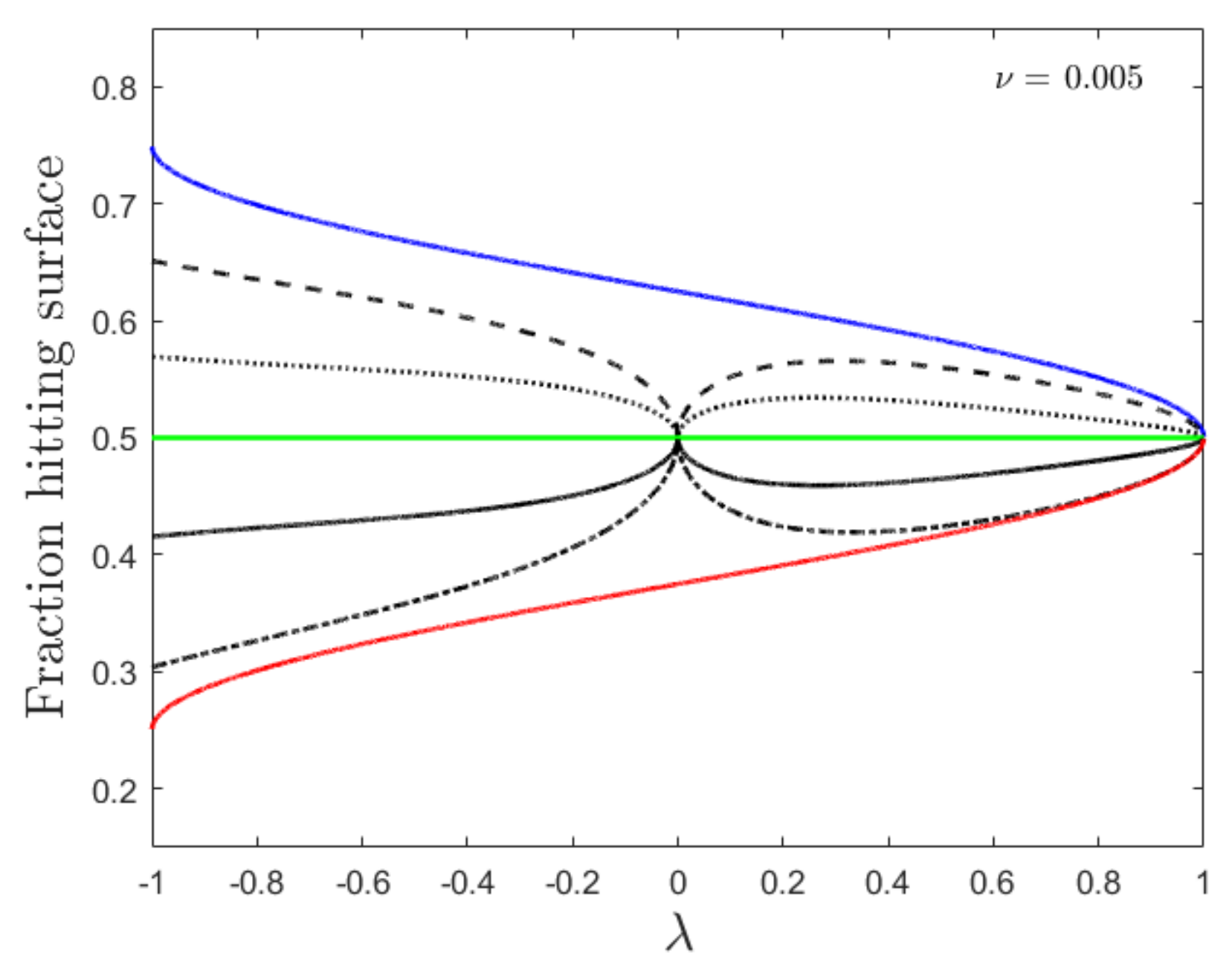}}
  \subcaptionbox{\label{fig:FHS_3depth_nu_0pt001}}{%
    \includegraphics[width=0.49\textwidth]{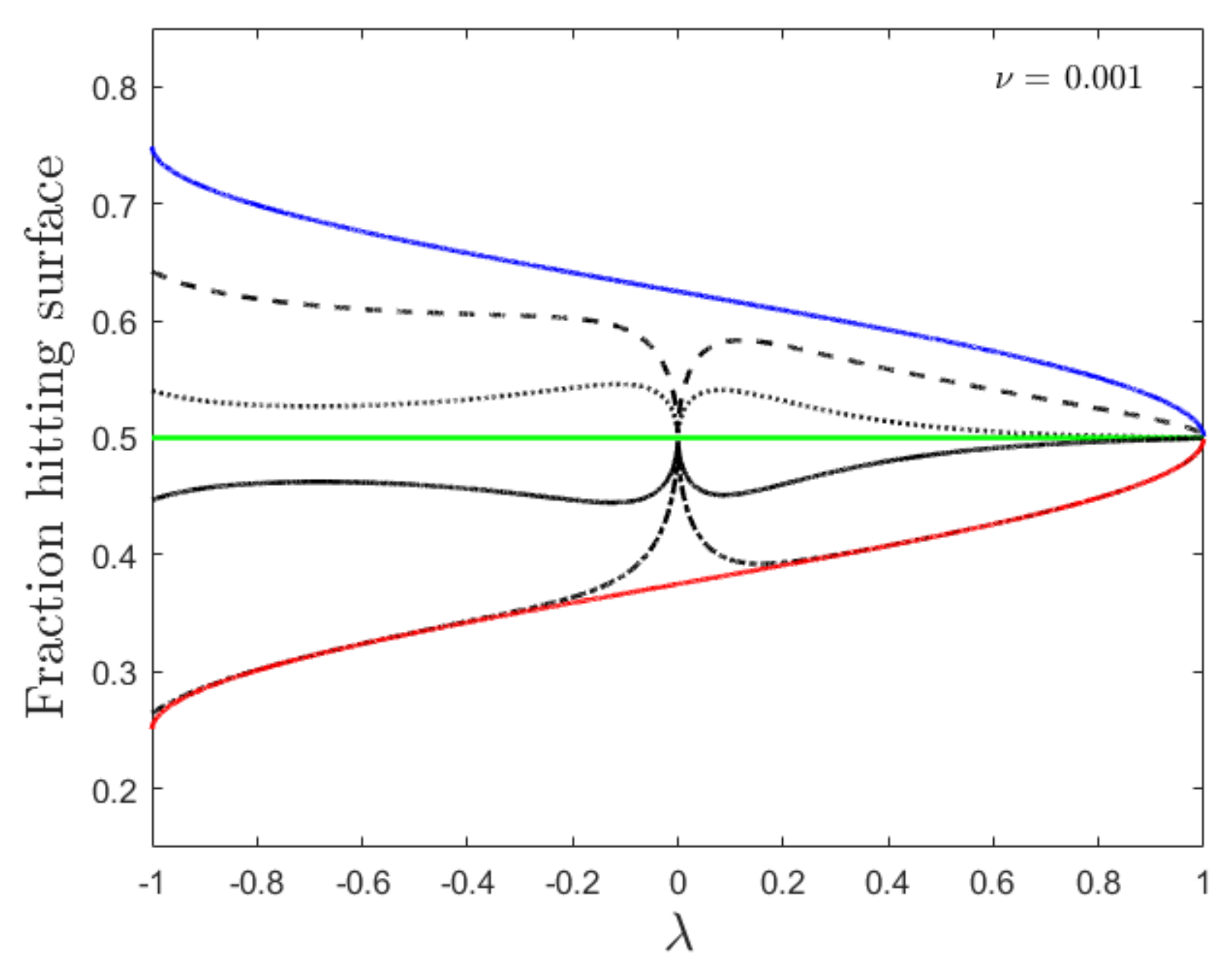}}
  \subcaptionbox{\label{fig:FHS_3depth_nu_0pt0001}}{%
    \includegraphics[width=0.49\textwidth]{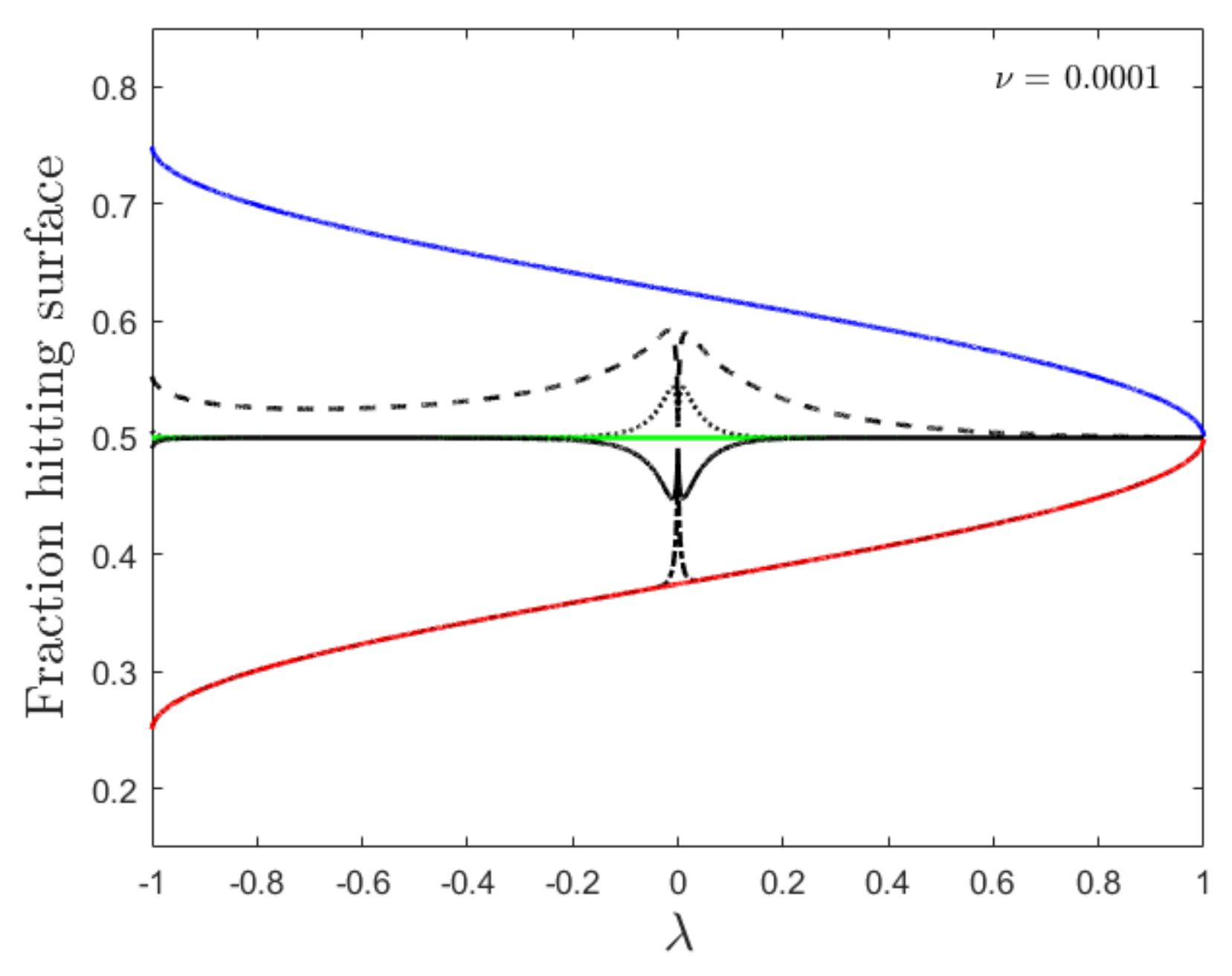}}
  \end{center}
  \caption{The fraction (Eq.~\eqref{eq:FHS}) of microswimmers that hit the surface as a function of shape and starting depth for different swimming speeds. Black curves give the fraction hitting surface based on Eq.~\eqref{eq:FHS}, computed using Eqs.~\eqref{eq:boundingcurve_left_positivelambda}--\eqref{eq:boundingcurve_right_negativelambda}. Blue and red curves are upper (Eq.~\eqref{eq:FHS_upper_bound}) and lower bounding (Eq.~\eqref{eq:FHS_lower_bound}) curves, respectively.}
  \label{fig:FHS_changing_nu}
\end{figure}

Fig.~\ref{fig:FHS_changing_nu} shows the variation of FHS with shape for different swimming speeds. (Recall that $\nu$ is a small parameter such that $\nu \ll \alpha$.) As $\nu$ decreases, FHS tends towards a value of 0.5 for all shapes and starting depths, with prolate swimmers ($\lambda>0$) achieving this value earlier than oblate swimmers ($\lambda<0$). Disk-shaped swimmers ($\lambda = -1$) and near-spherical shapes ($-0.2 \lesssim \lambda \lesssim 0.2$) are the last to achieve an FHS value of 0.5 as swimming speed decreases further. As before, exactly half of the spherical swimmers ($\lambda = 0$) and fiber-shaped swimmers ($\lambda=1$) hit the surface and exactly half of the swimmers of all shapes starting at the special depth $Z_0 = -\log{\sqrt{2}}$ hit the surface.

It is possible to find upper and lower bounds for FHS in the limit of small swimming speed for very shallow starting depths ($Z_0 \rightarrow 0$) and very deep starting depths ($Z_0 \rightarrow -\infty$), respectively, by setting $\nu \rightarrow 0$ in Eq.~\eqref{eq:Delta}. Since the left and right bounding curves for swimmers that end up at infinite depth are calculated by setting $Z_0 \rightarrow -\infty$ and $Z_0 \rightarrow 0$ in Eqs.~\eqref{eq:boundingcurve_left_positivelambda}--\eqref{eq:boundingcurve_right_negativelambda}, it is important to take the appropriate $Z_0$ limit before taking the $\nu \rightarrow 0$ limit. Setting $Z_0 \rightarrow 0$ and then $\nu \rightarrow 0$ gives $\Delta(\ee^{2Z_0}) \rightarrow 1$ and $\Delta(1-\ee^{2Z_0}) \rightarrow 0$ in Eqs.~\eqref{eq:boundingcurve_left_positivelambda}--\eqref{eq:boundingcurve_right_negativelambda}, resulting in the upper bounding curve
\begin{equation}
  \mathrm{FHS} \le
    \mathrm{FHS}^{(\text{upper})} = \frac{3}{4} - \frac{\Lambda}{2\pi}.
    \label{eq:FHS_upper_bound}
\end{equation}
\smallskip
Similarly, setting $Z_0\rightarrow -\infty$ and then $\nu \rightarrow 0$ gives $\Delta(\ee^{2Z_0})\rightarrow 0$ and $\Delta(1-\ee^{2Z_0}) \rightarrow 1$ in Eqs.~\eqref{eq:boundingcurve_left_positivelambda}--\eqref{eq:boundingcurve_right_negativelambda}, resulting in the lower bounding curve
\begin{equation}
  \mathrm{FHS} \ge
    \mathrm{FHS}^{(\text{lower})} = \frac{1}{4} + \frac{\Lambda}{2\pi}.
    \label{eq:FHS_lower_bound}
\end{equation}
Figure~\ref{fig:FHS_changing_nu} shows that Eqs.~\eqref{eq:FHS_upper_bound} and~\eqref{eq:FHS_lower_bound} provide upper and lower bounds on FHS that are symmetric about 0.5.  \change{The bounds are always worse near~$\lambda=0$, since in the limit $\lambda\rightarrow0$ (spherical swimmer) the value of~$x$ is irrelevant in Eq.~\eqref{eq:Delta}.}

\section{Discussion\label{sec:discuss}}
By using a two-timescale expansion, we have derived wave-averaged equations of motion for microswimmers in surface waves. The wave-averaged system reveals several aspects of microswimmer motion: (1) there are critical angles of microswimmer orientation that correspond to wave-induced preferred orientations, which are independent of swimming speed and only functions of microswimmer shape; (2) the horizontal and vertical motions are decoupled, with only the vertical component being coupled to the rotational motion; and (3) microswimmer trajectories are not too sensitive to the precise wave phase of the initial conditions, except when the initial orientation is near the critical angles.

By deriving the trajectories of microswimmers in the phase space spanned by the vertical position and orientation, we have shown that the vertical transport of microswimmers in surface waves is dependent on shape and swimming speed. In particular, microswimmers that begin with a component of their swimming velocity in the downward direction can still return to the surface as a result of flow-induced reorientation and it is possible for a microswimmer to swim arbitrarily deep and still return to the surface (though this is a very slow process). In general, this return to the surface is a function of microswimmer shape, starting depth, and swimming speed. The fraction of initially randomly oriented swimmers that return to the surface monotonically decreases with starting depth, with a special depth at which exactly half the swimmers return to the surface irrespective of shape or swimming speed. As a function of shape, there is more variability in the fraction hitting the surface for oblate microswimmers compared to prolate ones. For spherical swimmers and fiber-shaped (infinitely prolate) swimmers, exactly half return to the surface irrespective of starting depth and swimming speed. As a function of swimming speed, the fraction hitting the surface approaches a value of one half for all shapes as the swimming speed decreases. We also calculate upper and lower bounds on this fraction for small swimming speeds. This is a useful limit since we recall that microswimmer swimming speeds are at least an order of magnitude smaller than the typical wave-orbital fluid velocities. Here, we see that oblate microswimmers are more likely to reach the surface when they are already near the surface and more likely to dive to infinite depth when they are already at quite deep. For prolate swimmers, the likelihoods of reaching the surface or diving to large depths are less extreme. This suggests that prolate swimmers are more likely to achieve vertical migration in the desired direction with active changes in swimming behaviour than oblate swimmers.

Finally, we note that while our microswimmer model neglects many complexities of natural and artificial swimmers (\textit{e.g.}, non-axisymmetric shapes, differences in swimmer density with respect to the fluid, non-uniform mass distributions), the results clearly show that shape and swimming speed are important aspects for how microswimmers navigate aquatic environments in the presence of surface waves. Our model is also restricted to purely two-dimensional (2D) dynamics where the microswimmer axis is restricted to the flow plane. \change{However, in Appendix~\ref{sec:3D}, an analysis of the three-dimensional (3D) system suggests that the in-plane (2D) motion captures the essential features of the 3D system, except near the free surface}. We anticipate that future work which considers additional effects, such as the presence of noise (\textit{e.g.}, rotational diffusion), bottom-heaviness (\textit{e.g.}, gyrotaxis), \change{bias towards light (\textit{e.g.}, phototaxis)}, and buoyancy (\textit{e.g.}, settling), would modify the current results and provide an even more realistic picture of how plankton navigate wavy flow near the ocean surface.

\bibliography{swimwavePRF}

\appendix

\section{Three-dimensional model}
\label{sec:3D}

In this appendix we describe the full three-dimensional model for a spheroidal swimmer, and its two-time reduction.  In three dimensions the swimmer position has coordinates~$(x,y,z)$, where~$y$ is perpendicular to the waves.  We use polar angles $\phi$ and~$\theta$ to define swimming direction
\begin{equation}
  p_{x} = \sin\phi\,\sin\theta, \qquad p_{y} = \cos\theta,
  \qquad p_{z} = \cos\phi\,\sin\theta
\end{equation}
which reduces to Eq.~\eqref{eq:swimdir} for~$\theta=\pi/2$, where the angle~$\theta \in [0,\pi]$ is measured from the $y$ axis.  Jeffery's equations~\eqref{eq:Jefferys} generalized to three dimensions are
\begin{subequations}
\label{eq:Jefferys_3D}
\begin{align}
    \lambda^{-1}\,\dot{p}_x &= S_{xx}p_{x}(1-(p_{x}^2-p_{z}^2))+S_{xz}p_{z}(1-2p_{x}^2) \\
    \lambda^{-1}\,\dot{p}_y &=
    -p_{y}[S_{xx}(p_{x}^2-p_{z}^2)+2S_{xz}p_{x}p_{z}] \\
    \lambda^{-1}\,\dot{p}_z &=
    S_{xz}p_{x}(1-2p_{z}^2)-S_{xx}p_{z}(1+(p_{x}^2-p_{z}^2)),
\end{align}
\end{subequations}
which can be written in terms of angles as
\begin{subequations}
\label{eq:Jefferys_reduced_3D}
\begin{align}
  \lambda^{-1}\dot{\phi} &= S_{xx}\sin{2\phi} + S_{xz}\cos{2\phi} \\
  \lambda^{-1}\dot{\theta} &=
  \tfrac12\sin2\theta\,
  \l[S_{xz}\sin{2\phi} - S_{xx}\cos{2\phi}\r] .
\end{align}
\end{subequations}
With these extended coordinates and angles, the nondimensional ODEs~\eqref{eq:ODEs_dimensionless} generalize in three dimensions to
\begin{subequations}
  \label{eq:ODEs_dimensionless_3D}
  \begin{align}
    \dot{x} &= \alpha\,\ee^{z}\cos{(x - t)}
    + \nu\sin\phi\,\sin\theta \\
    \dot{y} &= \nu\cos\theta \\
    \dot{z} &= \alpha\,\ee^{z}\sin{(x - t)}
    + \nu\cos\phi\,\sin\theta \\
    \dot{\phi} &= \lambda \alpha\,\ee^{z}
    \cos(x -t + 2\phi)\\
    \dot{\theta} &= \tfrac12\lambda \alpha \ee^{z}\sin2\theta\,
    \sin(x - t + 2\phi).
  \end{align}
\end{subequations}
We introduce a fast and a slow time scale as in Eq.~\eqref{eq:twotime}, and rescale~$\alpha \rightarrow \eps\,\alpha$, $\nu \rightarrow \eps^2\,\nu$. After expanding all the quantities in powers of~$\eps$, we can carry out a similar procedure as in Section~\ref{sec:twotime} to find
%\begin{equation}
%  x_0 = X(T), \qquad y_0 = Y(T), \qquad z_0 = Z(T),\qquad \phi_0 = \Phi(T),
%  \qquad \theta_0 = \Theta(T)
%  \label{eq:slowvar_3D}
%\end{equation}
the 3D version of Eq.~\eqref{eq:ODEs_dimensionless_eps2_solv}:
\begin{subequations}
\label{eq:ODEs_dimensionless_kappainf_eps2_solv}
\begin{align}
  \partial_T X &= \nu\sin\Phi\sin\Theta + \alpha^2\,\ee^{2Z} \\
  \partial_T Y  &= \nu\cos\Theta \\
  \partial_T Z  &= \nu\cos\Phi\sin\Theta
  \label{eq:ODEs_dimensionless_kappainf_eps2_solv_dTZ}
  \\
  \partial_T \Phi  &=  \lambda \alpha^2\,\ee^{2Z} \l(\lambda + \cos2\Phi\r)
  \label{eq:ODEs_dimensionless_kappainf_eps2_solv_dTphi}
  \\
  \partial_T \Theta  &=  \tfrac12\lambda \alpha^2\,\ee^{2Z} \sin2\Theta\sin2\Phi
  \label{eq:ODEs_dimensionless_kappainf_eps2_solv_dTtheta}
\end{align}
\end{subequations}
where capital letters denote averaged (slow) variables.  Unlike the system~\eqref{eq:ODEs_dimensionless_eps2_solv}, for~\eqref{eq:ODEs_dimensionless_kappainf_eps2_solv} we can no longer solve the $Z$--$\Phi$ equations by themselves: we must instead solve the coupled $Z$--$\Phi$--$\Theta$ subsystem, which is harder.  The~$\sin2\Theta$ factor on the right-hand side of~\eqref{eq:ODEs_dimensionless_kappainf_eps2_solv_dTtheta} implies that~$\Theta \in [0,\pi]$ never crosses~$\Theta=\pi/2$.  Hence, throughout the time evolution we have~$\sin\Theta \ge 0$, and~$\cos\Theta$ has the same sign as~$\cos\Theta_0$.

Some progress can be made by dividing~\eqref{eq:ODEs_dimensionless_kappainf_eps2_solv_dTphi} by~\eqref{eq:ODEs_dimensionless_kappainf_eps2_solv_dTtheta}, thus eliminating~$Z$:
\begin{equation}
    \label{PhiTheta_sol_step1}
    \frac{d{\Phi}}{d{\Theta}}
    =
    \frac{2\l(\lambda+\cos{2\Phi}\r)}{\sin{2\Theta}\sin{2\Phi}}\,.
\end{equation}
This can be solved as
\begin{equation}
    \l|\frac{\lambda+\cos{2\Phi}}{\lambda+\cos{2\Phi_0}}\r|^{1/2}=\frac{\cot{\Theta}}{\cot{\Theta_0}}
    \label{eq:3DThetaPhi_relationship}
\end{equation}
where~$\Phi_0$ and~$\Theta_0$ are initial conditions.
\newcommand{\CC}{C}%
%We define a constant
%\begin{equation}
%    \CC = \frac{\cot{\Theta_0}}{\l|\lambda+\cos{2\Phi_0}\r|^{1/2}}
%\end{equation}
We can rewrite~\eqref{eq:3DThetaPhi_relationship} in terms of $\sin\Theta$,
\begin{equation}
  \sin{\Theta}
  =
  \l(1 + \cot^2\Theta_0\,
  \frac{\l|\lambda + \cos{2\Phi}\r|}{\l|\lambda + \cos{2\Phi_0}\r|}
  \r)^{-1/2}
  %=
  %\frac{1}{\sqrt{1 + \CC^2\l|\lambda + \cos{2\Phi}\r|}}
  \label{eq:sinTheta}
\end{equation}
where we took the `+' solution for the square root, since~$\sin\Theta\ge0$.

\begin{figure}
  \centering
  \includegraphics[width=0.6\textwidth]{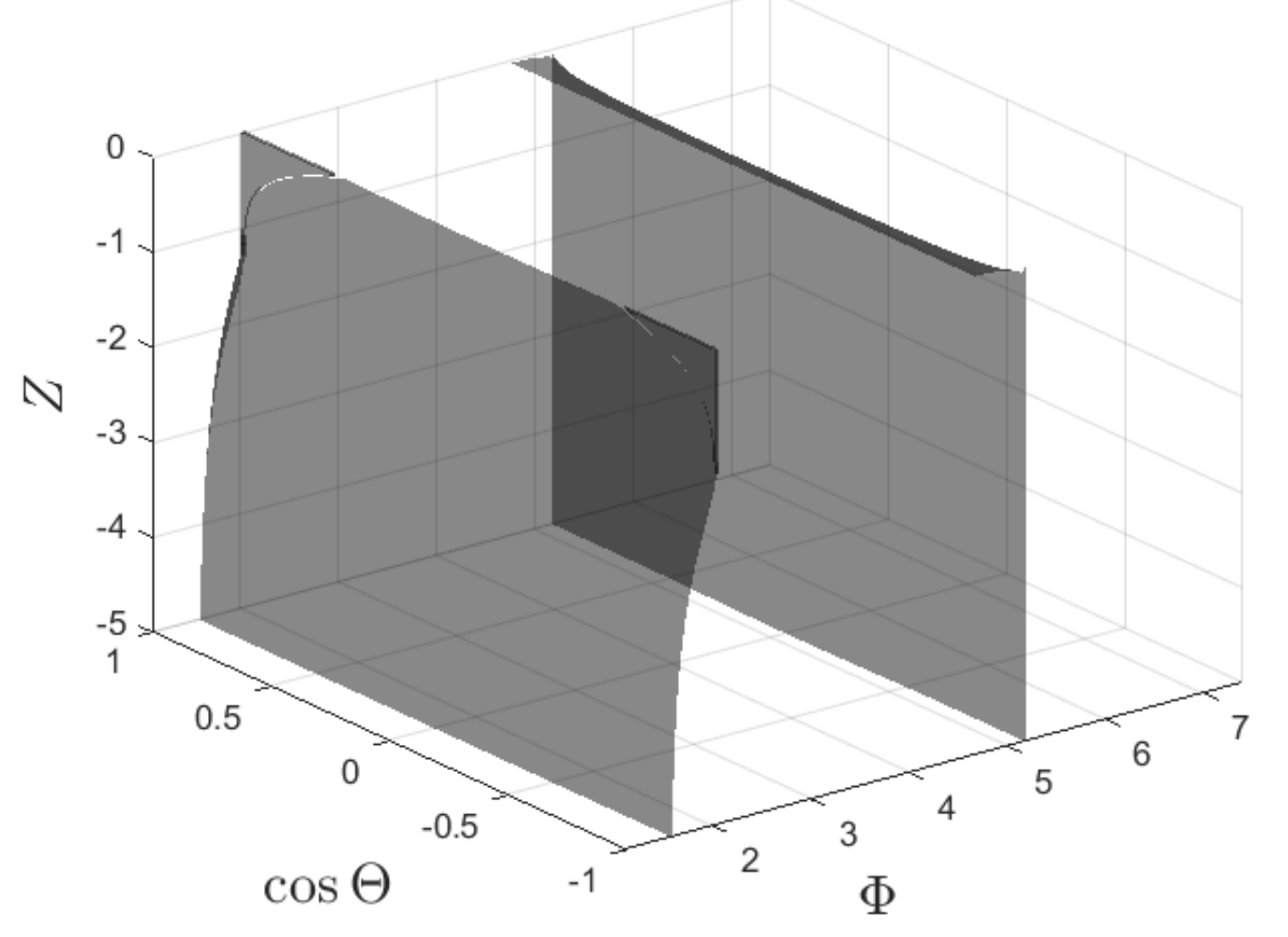}
  \caption{The three-dimensional region boundaries for~$\alpha=0.1$,
    $\nu=0.005$, and $\lambda=0.6$.  Initial conditions between the two
    surfaces for~$\cos\Theta_0=0$ correspond to the gray region in
    Fig.~\ref{fig:phaseportrait_Z_Phi_positive}.  The two bounding surfaces
    are relatively independent of~$\Theta_0$, except near $Z_0=0$.}
  \label{fig:3D_decision_boundary_CosTheta}
\end{figure}

We've solved for~$\sin\Theta$ as a function of~$\Phi$, so we can now solve the~$Z$--$\Phi$ system in~\eqref{eq:ODEs_dimensionless_kappainf_eps2_solv}.  Diviving~\eqref{eq:ODEs_dimensionless_kappainf_eps2_solv_dTZ} by~\eqref{eq:ODEs_dimensionless_kappainf_eps2_solv_dTphi}, we obtain an ODE that generalizes \eqref{eq:dexp2ZdsinPhi} to 3D:
\begin{equation}
    \frac{d\l(\ee^{2Z}\r)}{d\l(\sin{\Phi}\r)}
    =
    \frac{\nu}{\lambda\alpha^2}\,
    \frac{1}{\sin^2{\Lambda-\sin^2{\Phi}}}\,
    \sin\Theta
    %{\l(1 + 2\CC^2\l|\sin^2{\Lambda}-\sin^2{\Phi}\r|\r)^{-1/2}}
  \label{eq:dexp2ZdsinPhi_3D}
\end{equation}
where $\sin\Theta$ stands for the expression~\eqref{eq:sinTheta}.  We recover~\eqref{eq:dexp2ZdsinPhi} by setting~$\Theta=\Theta_0=\pi/2$.  Luckily, we can still solve~\eqref{eq:dexp2ZdsinPhi_3D} to obtain a solution analogous to~\eqref{eq:exp2Zsol}:
\begin{align}
   \ee^{2Z}-\ee^{2Z_0}
   =
   \frac{\nu}{2\lambda\alpha^2\sin{\Lambda}}
   \log{\l(\frac{\sin{\Lambda} + \sin{\Phi}\sin\Theta}{\sin{\Lambda} - \sin{\Phi}\sin\Theta}
     \,\frac{\sin{\Lambda} - \sin{\Phi_0}\sin\Theta_0}{\sin{\Lambda} + \sin{\Phi_0}\sin\Theta_0}\r)}.
   \label{eq:exp2Zsol_3D}
\end{align}
With the definition~\eqref{eq:Delta} for~$\Delta(x)$,
%we can rewrite this as
%\begin{equation}
%  \Delta\l(\ee^{2Z}-\ee^{2Z_0}\r)
%  = \frac{\sin{\Lambda}\,\l(\sin{\Phi}\sin\Theta - \sin{\Phi_0}\sin\Theta_0\r)}%{\sin^2{\Lambda} - \sin{\Phi}\sin{\Phi_0}\sin\Theta\sin\Theta_0}\,.
%\end{equation}
we find an expression that generalizes~\eqref{eq:sinPhi0sol} to three dimensions:
\begin{equation}
  \sin\Phi_0
  =
  \frac{\sin\Phi\sin\Theta - \sin\Lambda\,\Delta(\ee^{2Z} - \ee^{2Z_0})}
  {\sin\Lambda - \sin\Phi\sin\Theta\,\Delta(\ee^{2Z} - \ee^{2Z_0})}
  \,\frac{\sin\Lambda}{\sin\Theta_0}\,.
  \label{eq:sinPhi0sol_3D}
\end{equation}
We can now analyze the fraction of swimmers hitting the surface as we did in Section~\ref{sec:probhit}.  This thorny analysis can be carried out, but we have found in practice that this changes our earlier conclusions very little (see Figs.~\ref{fig:fraction_hitting_surface_2D_3D} and~\ref{fig:3D_decision_boundary_CosTheta}).  The reason is that~$\Phi$ tends to asymptote to critical angles where~$\lambda + \cos2\Phi = 0$, so that $\sin\Theta$ converges to~$1$ according to~\eqref{eq:sinTheta}, recovering the two-dimensional limit.

\end{document}